\documentclass[aps,pra,onecolumn]{revtex4}
\usepackage{graphicx}
\usepackage{hyperref}

\begin{document}
\title{Efficient retrieval of a single excitation stored in an atomic ensemble}
\author{Julien Laurat, Hugues de Riedmatten, Daniel Felinto, Chin-Wen Chou, Erik W. Schomburg, and H. Jeff Kimble}\address{Norman Bridge Laboratory of Physics 12-33, California
Institute of Technology, Pasadena, California 91125, USA}

\begin{abstract}
We report significant improvements in the retrieval efficiency of
a single excitation stored in an atomic ensemble and in the
subsequent generation of strongly correlated pairs of photons. A
50$\%$ probability of transforming the stored excitation into one
photon in a well-defined spatio-temporal mode at the output of the
ensemble is demonstrated. These improvements are illustrated by
the generation of high-quality heralded single photons with a
suppression of the two-photon component below 1$\%$ of the value
for a coherent state. A broad characterization of our system is
performed for different parameters in order to provide input for
the future design of realistic quantum networks.
\end{abstract}
\maketitle

\begin{figure}[htpb!]
\begin{center}
\includegraphics[width=11cm]{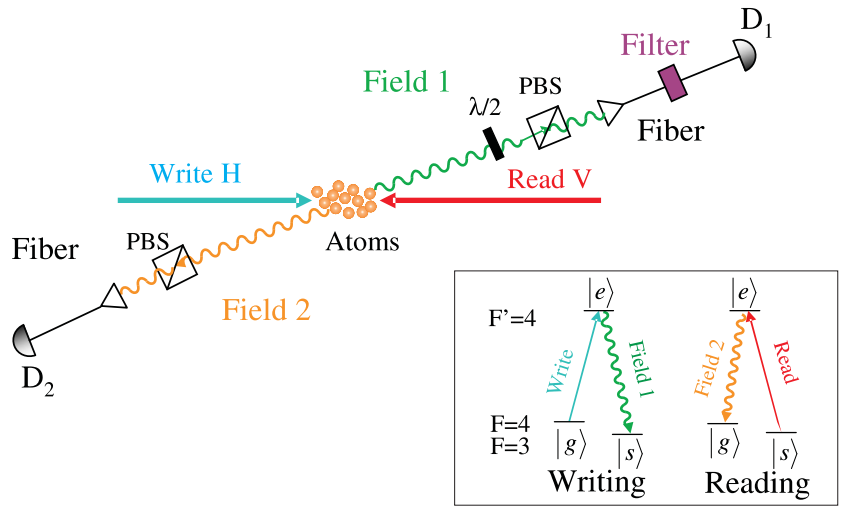}
\end{center}
\caption{\label{setup}Experimental setup. PBS stands for
polarizing beam splitter, H and V respectively for horizontal and
vertical polarization. The Filter stage for field 1 corresponds to
a paraffin-coated vapor cell (see text for details). The inset
illustrates the relevant atomic level scheme.}
\end{figure}

A basic requirement for long distance quantum communication is the
ability to efficiently interface atoms and light. Deterministic or
heralded storage of light in atomic systems is essential for
guaranteeing the scalability of protocols to distribute quantum
entanglement over large distances, such as in the quantum repeater
scheme \cite{rep}. In 2001, a significant step towards the
realization of a quantum repeater was the proposal by Duan, Lukin,
Cirac, and Zoller (DLCZ) of an alternative design involving atomic
ensembles, linear optics, and single photon detectors \cite{DLCZ}.
The building block of this roadmap is a large ensemble of
identical atoms with a $\Lambda$-type level configuration as
sketched in Figure \ref{setup}. A weak write pulse induces
spontaneous Raman scattering of a photon in field 1, transferring
an atom in the ensemble to the initially empty $|s\rangle$ ground
state. For a low enough write power, such that two excitations to
the $|s\rangle$ ground state are unlikely to occur, the detection
of the field-1 photon heralds the storage of a single spin
excitation distributed among the whole ensemble. A classical read
pulse can later, after a user-defined delay, transfer this atomic
excitation into another photonic mode (field 2). These scattering
events are collectively enhanced thanks to a many-atom
interference effect and can result in a high signal-to-noise ratio
\cite{PRADLCZ}. By following this line, nonclassical correlations
\cite{KuzmichNature1,angle,harris,PRAFelinto} and entanglement
\cite{Matsu1} have been observed between pairs of photons emitted
by a single atomic ensemble. By combining the output of two
different ensembles, as originally suggested in the DLCZ protocol,
heralded entanglement between two remote ensembles has been
recently demonstrated \cite{ChouNature}, paving the way for more
complex implementations of DLCZ schemes. \textit{A posteriori}
(probabilistic) polarization entanglement between two distant
ensembles has also been demonstrated recently \cite{Matsu2}, which
does not lead to scalable capabilities for quantum networks.

However, up to now, experiments have been plagued by low retrieval
efficiency of the stored excitation and background noise. The low
efficiency limits the extension to more complicated memory-based
protocols, as their success relies on many-fold coincidences.
Moreover, background noise prevents prior experiments from
reaching the very low excitation regime, in which high photon-pair
correlation and more pure conditional states can be obtained. We
report here significant experimental progress to overcome these
two limitations, which is crucial to enable realistic quantum
networks. As an illustration, high-quality heralded photons are
generated. These conditional single photons are spectrally narrow
and emitted in a well-defined spatio-temporal mode. They are thus
well-suited for diverse interactions between light and matter
\cite{Chaneliere,lukin,QI}.

The experimental setup is shown in Figure \ref{setup}. The
optically thick atomic ensemble is obtained from cold cesium atoms
in a magneto-optical trap (MOT). The Cs hyperfine levels
$\{|6S_{1/2}, F=4\rangle, |6S_{1/2}, F=3\rangle, |6P_{3/2},
F'=4\rangle\}$ correspond to levels $\{|g\rangle, |s\rangle,
|e\rangle\}$. The energy difference between the excited state and
the ground states corresponds to a wavelength of 852 nm. The
excitation and retrieval are carried out in a cyclic fashion. At a
frequency of 40 Hz, the magnetic field is switched off for 5 ms.
After waiting about 3 ms for the magnetic field to decay
\cite{PRAFelinto}, a sequence of 1100 trials with duration 2 $\mu
s$ begins. For each trial, the atoms are initially prepared in
level $|g\rangle$ by illuminating the cloud with trapping and
repumping light for 0.5 and 1.1 $\mu s$, respectively. A weak
write pulse, with a 200 $\mu m$ beam waist, and detuned 10 MHz
below the $|g\rangle$ to $|e\rangle$ transition, is first sent
into the atomic sample. For a low enough excitation probability,
the detection of a field-1 photon heralds the transfer of an atom
from the $|g\rangle$ to the $|s\rangle$ ground state. The read
pulse, which is on resonance with the $|s\rangle$ to $|e\rangle$
transition, orthogonally polarized with respect to the write beam
and mode-matched to it, is then fired after a programmable delay.
Both write and read pulses have about 30 ns duration. Fields 1 and
2 are directed into fibers with a $3^{\circ}$ angle relative to
the common direction defined by write and read beams
\cite{harris}, and with a waist of 50 $\mu m$ for the projected
mode in the atomic sample. This angle allows an efficient spatial
filtering. Field 1 (2) is detected, with a polarization orthogonal
to the write (read) beam, by single-photon silicon avalanche
photodiodes (APD), and the electronic signals are sent to a data
acquisition card, in order to record the detection events and
analyze the correlations. At the end of the 5 ms sequence, a new
MOT is formed. Note that, before detection, field 1 passes through
a filtering stage, in which it goes out of the fiber, through a
paraffin-coated Cs vapor cell, and back into the fiber
\cite{KuzmichNature1}. The vapor cell is initially prepared with
all atoms in $|g\rangle$. It filters out the photons in field 1
that are spontaneously emitted when the atoms in the sample go
back to $|g\rangle$, which do not trigger the creation of the
desired collective state.

\begin{figure}[htpb!]
\begin{center}
\includegraphics[width=12cm]{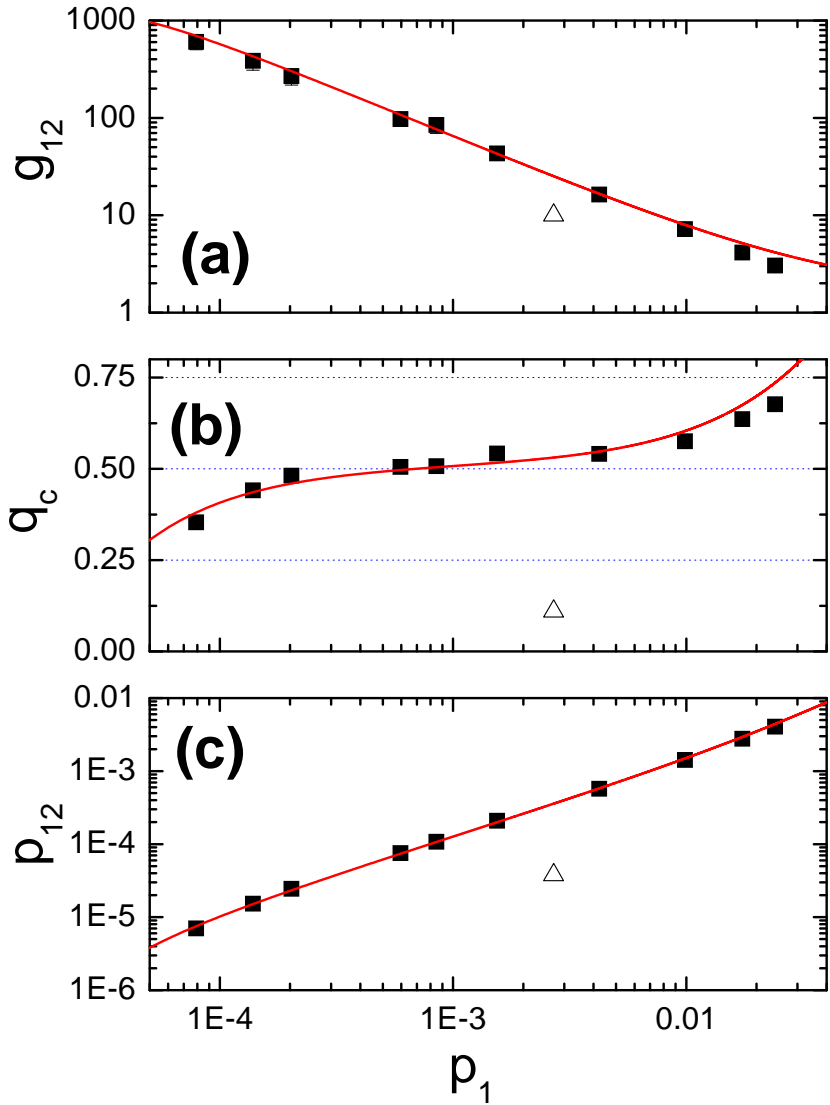}
\end{center}
\caption{\label{characterization}Characterization of the system as
a function of the probability $p_1$ to detect a photon from field
1. The frames give respectively the normalized intensity
cross-correlation function $g_{12}$ between fields 1 and 2, the
conditional retrieval efficiency $q_{c}$, and $p_{12}$, which is
the probability to detect a pair of photons, one in each field.
The parameter $q_c$ is obtained from the measured conditional
probability $p_c$ to detect field 2 once an event in field 1 has
been recorded by the relation $p_c=\eta q_c$. The two points with
lowest $p_1$ were obtained without trapping light (just repumper)
between trials, in order to reduce the noise level. All error bars
are smaller than the symbol size. The solid lines are from the
simplified model of Ref. \cite{ChouPRL}, with one set of
parameters for all plots. The triangle indicates the experimental
conditions for the entanglement measurement reported in Ref.
\cite{ChouNature}.}
\end{figure}

An important figure of merit in studying the correlations in a
photon-pair experiment is the normalized intensity
cross-correlation function $g_{12}=p_{12}/(p_1 p_2)$, where
$p_{12}$ is the probability to detect a pair of photons, and $p_i$
are the probabilities of detecting a single photon in field $i$.
In our case, measuring a value of $g_{12}$ larger than 2 is a
strong indication of non-classical correlations
\cite{KuzmichNature1,PRAFelinto}. Another measure for the quality
of the pair generation process is the degree of suppression $w$ of
the two-photon component of the field obtained from the retrieval
of the collective excitation (field 2 conditioned on a detection
in field 1), when compared to a coherent state
\cite{grangier,ChouPRL}. Another critical parameter to
characterize the efficiency of the setup, as underlined before, is
the ability to efficiently retrieve the stored excitation by
firing a read pulse. We will denote $q_c$ the probability to have
a photon in field 2, in a single spatial mode, at the output of
the atomic ensemble once an event has been recorded for field 1.
The conditional probability $p_c$ of having a detection event in
field 2 is, of course, lower due to experimental losses from the
atomic ensemble to the detector. If $\eta$ is the overall
detection efficiency for field 2, $q_c$ and $p_c$ are directly
related by $p_c=\eta q_c$. For our setup, $\eta=0.25$ due to
50$\%$ transmission loss in the field 2 pathway and the 50$\%$
detection efficiency of the APDs. The overall detection efficiency
for field 1 is also around 0.25.

Note that, for a given degree of correlation characterized by
$g_{12}$ and $w$, the actual operation of a quantum information
protocol, like the quantum repeater scheme proposed by DLCZ, can
be evaluated based on the values for $p_1$, $p_c$, and the memory
time of each site. A detailed analysis of the perspectives for
memory time in our system has been addressed in a previous paper
\cite{PRAFelinto}. In the present work, we are more concerned with
the other quantities, so that we fix our read/write delay at 300
ns.

Figure \ref{characterization} gives the experimental data for
$g_{12}$, the conditional retrieval efficiency $q_c$, and the
joint probability $p_{12}$, as functions of $p_1$. The rate of
coincidence counts per second can be obtained by multiplying
$p_{12}$ by the number of trials per second (44000 in our case).
Note that all measurements reported in this paper are not
corrected for any background or dark noise. The experimental
points are fitted according to the model introduced in
\cite{ChouPRL}, which assumes that the total fields at the output
of the MOT consist of a two-mode squeezed state plus background
fields in coherent states (which change in proportion to the write
field, such as light scattered from the write laser or background
fluorescence from uncorrelated atoms that are also excited by the
write beam) and two incoherent backgrounds to account for
processes that do not depend on the write beam intensity (residual
light from the environment or the MOT, and dark noise of the
detectors). These results for $g_{12}$ show more than a ten-fold
improvement on previous reported experiments
\cite{KuzmichNature1,PRAFelinto,Matsu1,ChouNature,Matsu2,Chaneliere,lukin,ChouPRL}.
For the highest value $g_{12}=600\pm 100$, the write pulse
contains about $10^4$ photons.

The shape of the $q_c$ curve reveals different regimes for
photon-pair generation. For high $p_1$, $q_c$ decreases as $p_1$
is reduced, since the multi-excitation processes decrease with the
energy of the write pulse. Once we reach the single excitation
regime, $q_c$ follows a plateau, since each detected field-1
photon (with whatever small rate) can lead to the subsequent
detection of only one field-2 photon. In this regime, the
probabilities for multi-excitation are negligible. When $p_1$
approaches the noise floor, however, $q_c$ starts to decrease
again due to false field-1 detection induced by the noise. From
the plateau on the $q_c$ curve, it results that the retrieval
efficiency of a single excitation in our system is approximately
50 $\%$ (corresponding to a measured $p_c=12.5\%$). This value
represents a three to five-fold improvement with respect to
previous works \cite{ChouNature,Chaneliere,lukin}. We note that
Ref. \cite{vuletic} has also investigated single-photon generation
via the collective emission of an atomic ensemble, here coupled to
an optical cavity. However, no direct measure of the single-photon
character of the retrieved field 2 was given. Moreover, a
retrieval efficiency was determined by way of the ratio
$p_2$/$p_1$, which has, contrary to $p_c$, the drawback of being
dependent on the characterization and modelling of the
uncorrelated background fields. For instance, applying the same
criteria to our measurements leads to a ``retrieval efficiency" of
around 160$\%$.

\begin{figure}[htpb!]
\begin{center}
\includegraphics[width=11cm]{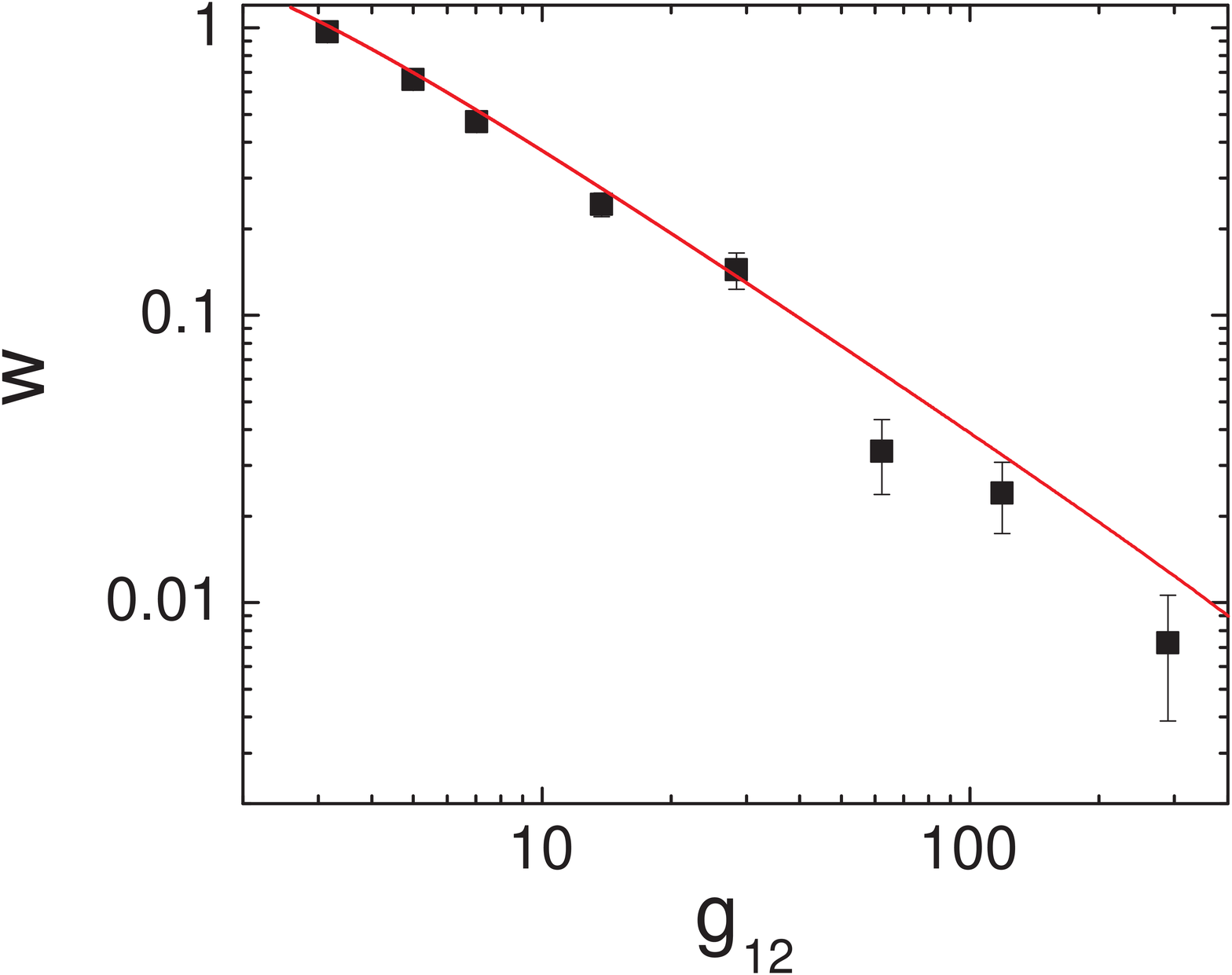}
\end{center}
\caption{\label{g2}Suppression of the two-photon component of
field 2 conditioned on the detection events in field 1, measured
by the $w$ parameter, as a function of $g_{12}$. The lowest value
obtained is $0.007\pm0.003$. All points in this plot were obtained
without trapping light between trials. The full curve is from a
fit based on the simplified model of Ref. \cite{ChouPRL}, with the
same parameters as in Figure 2.}
\end{figure}

As emphasized previously, the photon-pair correlations enable the
generation of heralded single photons. This method has been widely
and very successfully used over the past decades, firstly with
twin photons generated by an atomic cascade \cite{grangier}, then
by using the more efficient technique of parametric down
conversion \cite{hong}. More recently atomic ensembles were
introduced as an alternative source of single photons
\cite{Chaneliere,lukin,ChouPRL}. To assess the quality of the
single photon generated, a 50/50 fiber beam splitter (with 20\%
transmission loss) is inserted into the pathway of field 2, and
two detectors, $D_{2a}$ and $D_{2b}$, are used to record the
events. Figure \ref{g2} gives the parameter $w$ for the detection
of two photons from field 2, conditioned upon the detection of an
initial photon in field 1. This parameter measures the
single-photon character of field 2. For both experiment and
theory, this quantity was obtained from single and joint
probabilities by the expression \cite{grangier,ChouPRL}
\begin{equation}
w = \frac{p_1 p_{1,2a,2b}}{(p_{1,2a})(p_{1,2b})} \;,
\end{equation}
where $p_{1,2a,2b}$ indicates the probability for a triple
coincidence between the three detectors, and $p_{1,2a}$
($p_{1,2b}$) gives the probability for coincidences between
detectors $D_1$ and $D_{2a}$ ($D_{2b}$). A best value of
$w=0.007\pm0.003$ is obtained. This represents more than a
twenty-fold improvement on the first reported value
\cite{ChouPRL}, and ten-fold on the subsequent extensions
\cite{Chaneliere,lukin}. Furthermore, this value is close to the
best reported ones in parametric down conversion \cite{fasel,ren}.
This low value makes this source relevant in the realization of
linear-optics quantum computation \cite{KLM}, where two-photon
contamination must be minimized. Note again that these photons are
narrowband and thus well-suited for light-atom interfacing
\cite{Chaneliere,lukin,QI}.

These large improvements were obtained after an empirical
adjustment of the duration and energy ($10^7$ photons per pulse)
of the read pulse, and an increased optical depth ($OD=12$) of the
ensemble. We also incorporated several improvements previously
implemented by other groups, like the off-axis detection geometry
(introduced by S. Harris's group \cite{harris} and later optimized
by A. Kuzmich's group \cite{Matsu1}) and the use of large-waist
write and read beams \cite{axel}. We also found that it is
essential to continue using the frequency filter in field 1. The
lack of this filter in recent experiments that rely completely on
postselection \cite{Matsu1,Matsu2,Chaneliere} might be responsible
for an effective decrease of $p_c$ and the measured correlations.

In conclusion, we have reported the efficient generation of
photon-pairs from atomic ensembles, following the building block
of the DLCZ roadmap. Apart from unprecedented nonclassical
correlations in this configuration, we obtained high retrieval
efficiency of the stored excitation around 50$\%$, which is a
critical parameter for the realization of sophisticated quantum
networks and memory-based protocols. The importance of having high
$p_c$ and $q_c$ is clearly illustrated in Ref. \cite{ChouNature},
in which the small values of these quantities limits the ability
to infer the degree of entanglement between the two distant
ensembles.

\section*{Acknowledgments}
This work was supported by the Disruptive Technology Office of the
Department of National Intelligence and by the National Science
Foundation. J.L. acknowledges financial support from the European
Community (Marie Curie Fellowship), and D.F. from CNPq (Brazilian
agency).


\begin{thebibliography}{}
\bibitem{rep} H.-J. Briegel, W. D\"{u}r, J. I. Cirac, and P. Zoller, ``Quantum repeaters: the role of imperfect local operations in quantum communication,'' Phys. Rev. Lett. \textbf{81}, 005932  (1998).
\bibitem{DLCZ} L.-M. Duan, M. D. Lukin, J. I. Cirac, and P. Zoller, ``Long-distance quantum communication with atomic ensembles and linear optics,'' Nature \textbf{414}, 413-418 (2001).
\bibitem{PRADLCZ} L. -M. Duan, J. I. Cirac, and P. Zoller, ``Three-dimensional theory for interaction between atomic ensembles and free-space light,'' Phys. Rev. A \textbf{66}, 023818 (2002).
\bibitem{KuzmichNature1} A. Kuzmich, W. P. Bowen, A. D. Boozer, A. Boca, C. W. Chou, L.-M. Duan, and H. J. Kimble, ``Generation of nonclassical photon pairs for scalable quantum communication with atomic ensembles,'' Nature \textbf{423}, 731-734 (2003).
\bibitem{angle} M. D. Eisaman, L. Childress, A. Andr{\'e}, F. Massou, A. S. Zibrov, and M. D. Lukin, ``Shaping quantum pulses of light via coherent atomic memory,'' Phys. Rev. Lett. \textbf{93}, 233602 (2004).
\bibitem{harris} V. Bali{\'c}, D. A. Braje, P. Kolchin, G. Y. Yin, and S. E. Harris, ``Generation of paired photons with controllable waveforms,'' Phys. Rev. Lett. \textbf{94}, 183601 (2005).
\bibitem{PRAFelinto} D. Felinto, C. W. Chou, H. de Riedmatten, S. V. Polyakov, and H. J. Kimble, ``Control of decoherence in the generation of photon pairs from atomic ensembles,'' Phys. Rev. A \textbf{72}, 053809 (2005).
\bibitem{Matsu1} D. N. Matsukevich, T. Chaneli{\`e}re, M. Bhattacharya, S.-Y. Lan, S. D. Jenkins, T. A. B. Kennedy, and A. Kuzmich, ``Entanglement of a photon and a collective atomic excitation,'' Phys. Rev. Lett. \textbf{95}, 040405 (2005).
\bibitem{ChouNature} C. W. Chou, H. de Riedmatten, D. Felinto, S. V. Polyakov, S. J. van Enk, and H. J. Kimble, ``Measurement-induced entanglement for excitation stored in remote atomic ensembles,'' Nature \textbf{438}, 828-832 (2005).
\bibitem{Matsu2} D. N. Matsukevich, T. Chaneli\`{e}re, S. D. Jenkins, S.-Y. Lan, T. A. B. Kennedy, and A. Kuzmich, ``Entanglement of remote atomic qubits,'' Phys. Rev. Lett. \textbf{96}, 030405 (2006).
\bibitem{Chaneliere} T. Chaneli{\`e}re, D. N. Matsukevich, S. D. Jenkins, S.-Y. Lan, T. A. B. Kennedy, and A. Kuzmich, ``Storage and retrieval of single photons transmitted between remote quantum memories,'' Nature \textbf{438}, 833-836 (2005).
\bibitem{lukin} M. D. Eisaman, A. Andr{\'e}, F. Massou, M. Fleischhauer, A. S. Zibrov, and M. D. Lukin, ``Electromagnetically induced transparency with tunable single-photon pulses,'' Nature \textbf{438}, 837-841 (2005).
\bibitem{QI}  D. Bouwmeester, A. Ekert, and A. Zeilinger, ``The Physics of Quantum Information''(Springer-Verlag, Berlin, Germany, 2001)


\bibitem{grangier} P. Grangier, G. Roger, and A. Aspect, ``Experimental evidence for a photon anticorrelation effect on a beam splitter - A new light on single-photon interferences,'' Europhys. Lett. \textbf{1}, 173-179 (1986).
\bibitem{ChouPRL} C. W. Chou, S. V. Polyakov, A. Kuzmich, and H. J. Kimble, ``Single-photon generation from stored excitation in an atomic ensemble,'' Phys. Rev. Lett. \textbf{92}, 213601 (2004).

\bibitem{vuletic} A. T. Black, J. K. Thompson, and V. Vuleti\c{c}, ``On-demand superradiant conversion of atomic spin gratings into single photons with high efficiency,'' Phys. Rev. Lett. \textbf{95}, 133601 (2005).
\bibitem{hong} C. K. Hong and L. Mandel, ``Experimental realization of a localized one-photon state,'' Phys. Rev. Lett. \textbf{56}, 58-60 (1986).
\bibitem{fasel} S. Fasel, O. Alibart, S. Tanzili, P. Baldi, A. Beveratos, N. Gisin, and H. Zbinden, ``High-quality asynchronous heralded single-photon source at telecom wavelength,'' New J. Phys. \textbf{6}, 163 (2004).
\bibitem{ren} A. B. U'Ren, C. Silberhorn, J. L. Ball, K. Banaszek, and I. A. Walmsley, ``Characterization of the nonclassical nature of conditionally prepared single photons,'' Phys. Rev. A \textbf{72}, 021802(R) (2005).
\bibitem{KLM} E. Knill, R. Laflamme, and G. J. Milburn, ``A scheme for efficient quantum computation with linear optics,'' Nature \textbf{409}, 46-52 (2001).

\bibitem{axel} A. Andr\'{e}, ``Nonclassical states of light and atomic ensembles: generation and new applications,'' Thesis, Harvard University (2005).


\end{thebibliography}
\end{document}